\documentclass[twocolumn,amsmath,amssymb,prl,superscriptaddress,showpacs]{revtex4}
\usepackage{graphicx}
\usepackage{appendix}
\usepackage{bm}
\usepackage{braket}
\usepackage{color}

\begin{document}

\title{Self-organized electronic superlattices in layered materials}

\author{Carmine Ortix} 
\affiliation{Institute for Theoretical Solid State Physics, IFW Dresden, D01171 Dresden, Germany}
\author{Carlo Di Castro}
\affiliation{ISC-CNR and Dipartimento di Fisica, Universit\`a di Roma 
``La Sapienza'', Piazzale  Aldo Moro 2, 00185 Roma, Italy.}
\author{Jos\'e Lorenzana}
\affiliation{ISC-CNR and Dipartimento di Fisica, Universit\`a di Roma 
``La Sapienza'', Piazzale  Aldo Moro 2, 00185 Roma, Italy.}

\begin{abstract}
We show  that in layered systems with electronic phase separation
tendency, the long-range Coulomb interaction can drive the
spontaneous formation of unidirectional superlattices of electronic
charge in a completely homogeneous crystalline background. In this
self-organized electronic heterostructure,  the ratio among the
number of crystalline planes in the minority and majority electronic
phases corresponds to Farey fractions with the superlattice period controlled
by the background charge density and the frustrating Coulomb interaction strength. 
The phase diagram displays Arnold tongues 
obeying a modified Farey
tree hierarchy and a devil's staircase typical of systems with
frustration among different scales. 
We further discuss the competition of these electronic superlattices,
recently observed in iron-based superconductors 
and
mixed valence compounds,
with in-plane electronically
modulated phases.
\end{abstract}
\pacs{73.21.Cd, 71.10.Hf, 64.75.Jk, 64.75.-g}
\maketitle

\date{\today}

Domain pattern formation is a beautiful example of cooperative
behavior in a variety of systems with competing interactions on different
length scales \cite{seu95}. In strongly correlated electronic systems,
electronic charge modulated phases are center stage with complex
phenomena as colossal magnetoresistance \cite{bec02,zha02,lai10},
magnetoelectric effects \cite{gro12}, spin-Peierls-like behavior \cite{ohw01} and
high-temperature superconductivity in
cuprates \cite{cas95,eme93,low94,tra95,abb05,cha12,ghi12,com14,net14,lan02}
and iron-based superconductors \cite{char12,tex12,par09}.  
In most of these systems short-range electronic correlations can and do 
often drive a tendency towards 
phase separation in electron-rich and electron-poor phases.
Unless the counterions are mobile, the appearance of a macroscopically
electronic phase segregated state is prohibited by the
long-range part of the Coulomb interaction.    
Henceforth, the system compromises by stabilizing electronic microemulsions of the competing phases \cite{eme93,low94,cas95,nus99}
in the form of bubbles and stripes in two-dimensional (2D) systems \cite{ort06},
and layers, rods and droplets in three-dimensional (3D) ones \cite{ort08}. 
Irregularities
leading to fractal-like interfaces \cite{fat99} can be ascribed to a
non-negligible role of quenched disorder possibly stabilizing glassy
states.  

It has been theoretically established \cite{ort06} that the appearance of these charge textures 
is particularly favored in 2D systems and indeed this phenomenon has been observed in materials which share a strictly 2D or quasi-2D layered structure.
In layered systems, however, electronic phase separation tendencies can potentially lead 
to a phase segregated state consisting of alternating uniformly charged electron-rich and electron-poor crystalline planes \cite{fin08} (hereafter the 1/1 structure) recently observed in the mixed valence compound LuFe$_2$O$_4$ \cite{gro12}. 
The questions remain whether  
larger period 
phase segregated states 
of the same kind    
can be spontaneously stabilized and how the appearance of these electronic heterostructures competes with the formation of electronic microemulsions within the crystal planes.  
In this Letter, we 
show that 
when the frustrating Coulomb energy cost is small compared to the phase-separation energy gain, 
a layered system self-organizes in unidirectional superlattices of electronic charge whose period depends on the average charge. 
The ensuing phase diagram from the homogeneous phase to the inhomogeneous one 
is characterized by Arnold tongues \cite{sch05} obeying a modified Farey tree construction and a devil's staircase analogous to dynamical systems with competing frequencies \cite{bak82} and solid-state structures with frustration among different scales \cite{ohw01}.

To illustrate the emergence of electronically charged superlattices, we consider a layered system in which  
the
coarse grained electronic free energy can be expressed
phenomenologically as a Landau expansion in powers of an 
electronic charge density order parameter \cite{ort08}  $\phi_l ({\bf r})$,
\begin{eqnarray}
\label{eq:f0longrange}
{\cal F}_{\phi}&=& \sum_l \int \left[ \phi_l ({\bf r})^2 -1
\right]^2 + | \nabla \phi_l ({\bf r})|^2 \,  d {\bf r}  \\ 
& & 
 +\sum_{l,l^{\prime}} \dfrac{Q^2}{2} \int \dfrac{\left[\phi_{l}({\bf r})-\bar{\phi}\right]
  \left[\phi_{l^{\prime}}({\bf
      r^{\prime}})-\bar{\phi}\right]}{\sqrt{({\bf r-r^{\prime}})^2+ a_3^2 (l-l^{\prime})^2}} \, d{\bf r} \, d{\bf
  r^{\prime}}, \nonumber  
\end{eqnarray}
 where $l$ is a layer index while ${\bf r}$ is a two-dimensional
 intralayer position vector (measured in units of the 
 in-plane bare correlation length $\xi$). The first term has a double
 well form favoring phase separation in the absence of the Coulomb
 interaction.  
Energies are measured in units of the barrier height whereas densities are in units such that in each layer $\phi_l({\bf r}) = \pm 1$  represent  
 the ideal densities $n^0_{+}$, $n^0_{-}$  of the electron rich and electron poor phases.
 More precisely, the physical electron areal density per plane $l$  is 
 $n_l({\bf r}) = (n^0_{+} + n^0_{-})/2 + \phi_l ({\bf r}) \left(n^0_{+} - n^0_{-} \right)/2$. 
The second term in Eq.~\ref{eq:f0longrange}  
accounts for short range surface energy effects.  
We assume very weak short-range interlayer couplings, as it is the
case in many layered materials, and thus neglect the associated 
energy cost to create electronic domain
walls in the stacking direction.  The third term is the Coulomb
energy, frustrating macroscopic phase separation, with $a_3$ as the
spacing among the layers and  $Q^2$ parameterizing
the strength of Coulomb frustration with respect to the barrier
height of the double-well. Charge neutrality is guaranteed assuming a completely rigid ionic background of positive charge $-\bar{\phi}$ with $\sum_l \int \phi_l ({\bf r}) d {\bf r} =  \bar{\phi} \sum_l \int   d {\bf r} $.  Hereafter we will refer to $\bar{\phi}$ as the global density. 
The ideal electron rich and electron poor physical densities $n^0_{+,-}$, as well as
any other parameter in the model, are determined by control parameters
like pressure, temperature, doping, etc. For example 
we expect $Q^2\sim 1/(T^*-T)$ below a temperature $T^*$ where
thermodynamic phase separation 
would start in the absence of the stabilizing effect of the long-range
Coulomb interaction.

We first consider the 1/1 structure with charge distribution \cite{fin08},
$\phi_l = \bar{\phi} + \delta
\phi \,  \mathrm{e}^{i \pi l} $. The difference of free energy density
among this modulated phase and the homogeneous one [see Supplemental Material] takes the form
$\delta f= \alpha \left(Q, \bar{\phi}\right)  \delta \phi^2 + \beta \,
\delta \phi^4$  with  $\alpha \left( Q , \bar{\phi} \right)$ vanishing
along a Gaussian second order transition line, 
$Q_{G}^{1/1} = {2}\left[(1 - 3  {\bar\phi}^2)/\pi a_3 \right]^{1/2}$. 
As $Q$ approaches $Q_G^{1/1}$ from above, 
the homogeneous phase is thus unstable toward a charge modulated phase.
This second-order phase transition cannot survive in the  $Q
\rightarrow 0$ limit since it predicts a modulated phase for $\left| \bar{\phi} \right| < 1 / \sqrt{3}$, in disagreement with macroscopic phase separation  that instead secures an inhomogeneous phase in the larger range $\left| \bar{\phi} \right| < 1 $. 
The way out of this discrepancy consists in considering inhomogeneous states with the electronic charge density modulated again along the stacking direction but with larger period arrangements. 
\begin{figure}
\includegraphics[width=\columnwidth]{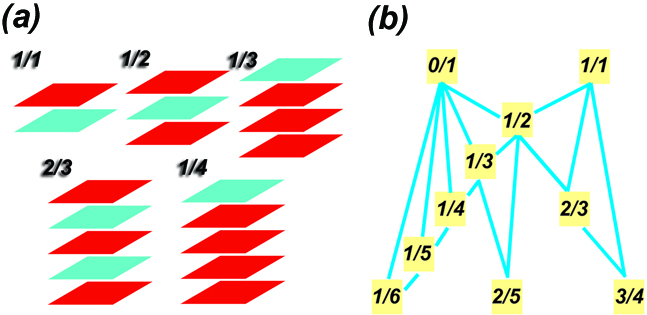}
\caption{(a) Sketch of the lowest period electronic heterostructures. (b)
Schematics of the modified Farey tree construction. Each fraction
$p/q$ is generated by its ancestors $p_a / q_a$ and $p_b / q_b$ as
$p/q=(p_a+p_b)/(q_a+q_b)$.}
\label{fig:fig1}
\end{figure}
These can be identified by setting $Q=0$ at first. The Maxwell construction predicts an  inhomogeneous state with ${\cal F}_{\phi} = 0$ where the density in the layers $\phi_l = \pm 1$ and the ratio between the 
total number of layers in the minority and the majority phase
is 
$\left( 1 - \left|\bar{\phi}\right| \right) / \left(1 + \left|\bar{\phi} \right| \right)$. 
In our model this does not uniquely fix the inhomogeneous state: due to the absence of stiffness in the staking
direction, 
there is a macroscopic degeneracy of different states fulfilling the
Maxwell construction. 
This degeneracy  is lifted once even infinitesimally small values of $Q$ are introduced. 
The long-range Coulomb interaction indeed selects the inhomogeneous states 
minimizing the periodicity  
while concomitantly maximizing the distance among the electronic domain walls within one period. 
This leads to uniquely defined superlattices of electronic charge [c.f. Fig.\ref{fig:fig1}(a) for a sketch of the lowest period arrangements]
where in one period made of $p+q$ layers the ratio among 
the number of layers ($p$) in the minority phase
and the number of layers ($q$) in the  majority phase realizes the series of Farey fractions 
$p/q = \left( 1 - \left|\bar{\phi}\right| \right) / \left(1 +
  \left|\bar{\phi} \right| \right)$ determined by changing the global
density from ${\bar \phi}\equiv \pm 1$ (the $0/1$ homogeneous phase) to ${\bar \phi}=0$ (the $1/1$ modulated phase). 

\begin{figure}
\includegraphics[width=\columnwidth]{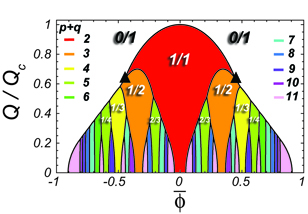}
\caption{ Phase diagram in the uniform density approximation for $a_3 \leq
\sqrt{6}$. The colors encode the periodicity $p+q$ of superlattices. 
The $\blacktriangle$ indicate the triple points 
separating a second-order transition line (above) from a first-order one (below). The strength of the Coulomb interaction has been measured in units of $Q_c=2 / \sqrt{\pi a_3}$.
The appearance of the larger periodicity tongues goes at the expense 
of the lower periodicity tongues which acquire a funnel shape. We have restricted to structures with $p+q \leq 11$. Allowing for arbitrary periodicities would shrink the lower part of the funnel to one point.}
\label{fig:fig1b}
\end{figure}

The complete devil's staircase with all possible rational values of $p/q$ obtained by varying the average density occurs only in the $Q \rightarrow 0$ limit.  
For finite $Q$, the electronically charged superlattices of largest lateral dimension  are progressively suppressed as $Q$ increases. 
To show this, we employ a uniform density approximation and assume that the charge density order parameters in the $p$ minority phase crystalline planes and the $q$ majority phase crystalline planes take constant values $\phi_{-}$ and $\phi_{+}$ respectively. The constraint of charge neutrality yields $p \phi_{-} + q \phi_{+} = (p+q) \bar{\phi}$. The finite Coulomb energy can be then expressed in terms of the electric field among two consecutive layers $E_l$ whose discontinuity at the $l$th layer is related to the local charge density by $E_l - E_{l-1} = 4 \pi \left(\phi_l-\bar{\phi}\right)$ where $\phi_l$  is uniquely determined by the spatial arrangement of the phase minority and phase majority layers in one period. The difference of free energy density among a modulated phase and the homogeneous one then takes the form 
\begin{eqnarray}
\delta f(p,q) &=& \dfrac{p}{q} \left[-2 + 6 \bar{\phi}^2 \right] \delta \phi_{-}^2 + 4 \,\bar{\phi}\, \dfrac{p}{q} \left(1 - \dfrac{p}{q} \right) \delta \phi_{-}^3 \label{eq:interUDA} \\ & & + \dfrac{p}{q} \left(1 - \dfrac{p}{q}+ \dfrac{p^2}{q^2} \right) \delta \phi_{-}^4+ \dfrac{1}{p+q} \sum_{l=1}^{p+q} \dfrac{Q^2 a_3}{8 \pi} E_l^2 \nonumber ,
\end{eqnarray}
where $p$ and $q$ are coprimes and $\delta \phi_{-}=\phi_{-}-\bar{\phi}$.
Fig.~\ref{fig:fig1b} shows the ensuing phase diagram in the $Q - \bar{\phi}$ plane obtained by minimizing Eq.~\ref{eq:interUDA} with respect to $\delta \phi_{-}$ and scanning for the most energetically favored modulated structure with 
maximum periodicity $p+q=11$.  
It shows the appearance of the so-called Arnold tongues \cite{sch05}
with the largest tongue appearing for the $p/q= 1/1$  superlattice. We
identify the hierarchy of Arnold tongues size to follow a modified
Farey tree construction: the tongues corresponding to $p/q$  rationals   $ \in  \left[ 0 , 1\right]$ appear by lowering $Q$ with increasing values of $p+q$  
(rather than $q$ alone as in the usual Farey tree construction) according to the rule that the largest tongue among $p/q$ and $p^{\prime} / q^{\prime}$ is $\left(p+p^{\prime}\right)/\left(q+q^{\prime}\right)$ [c.f. Fig.\ref{fig:fig1}(b) for the sequence of the modified Farey tree construction]. 
We emphasize that while, as discussed above, the transition from the homogeneous phase to the  1/1 modulated state is second-order, 
the onset of inhomogeneous structures with $p+q>2$ is  first-order thereby
leading to the appearance of a triple point in the phase diagram for the coexistence of the $0/1$, $1/1$ and $1/2$ phases   $\left\{ \bar{\phi}_T, Q_T \right\} = \left\{ 1/\sqrt{5} , \sqrt{8 / (5 \pi a_3)} \right\}$ determined by ${ min}_{\delta \phi_{-}} \delta f(1,1) \equiv { min}_{\delta \phi_{-}} \delta f(1,2) \equiv 0$. By decreasing the frustration strength additional triple points are found for the coexistence of the $0/1$, $1/q$, $1/(q+1)$ phases.

\begin{figure}
\includegraphics[width=\columnwidth]{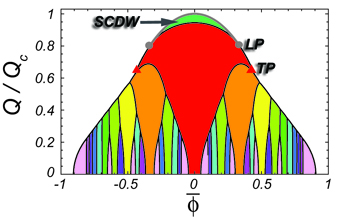}
\caption{ Phase diagram  including in-plane electronic microemulsions 
for a layered system with layer spacing $a_3= 3 \, \xi$. The strength of the Coulomb interaction has been measured in units of the maximum coupling strength $Q_c = 0.6623...$. The phase diagram displays two symmetric Lifshitz points (LP) and additional triple points (TP) at lower couplings.}
\label{fig:fig2}
\end{figure}

This, however, is not yet the end of the story. The Coulomb cost associated to the formation of a generic superlattice of electronic charge is indeed $\propto a_3$ [c.f. Eq.~\ref{eq:interUDA}] 
as it can be intuitively understood by considering two subsequent layers of the coexisting phases as a capacitance with $C \propto a_3^{-1}$ and Coulomb energy 
$E_{c} \propto C^{-1}$. 
It is then conceivable that by increasing 
the layer stacking distance the appearance of in-plane electronic microemulsions can preempt the formation of the electronic heterostructures discussed so far. This indeed occurs for $a_3 \geq \sqrt{6}$ [c.f. Supplemental Material for a detailed derivation]. 
To illustrate this point, we allow also for in-plane electronic charge density modulations. 
For a generic periodic texture, the electronic charge density  can be expanded in momentum space as 
$\phi_{l}({\bf r})=\overline{\phi}+\sum_{{\bf G}} \phi_{{\bf G} , l} \, \, 
 e^{i \, {\bf G} \cdot ({\bf r}+{\bf T}_l  )}$
where the ${\bf G}$'s are intralayer reciprocal lattice wavevectors
and the ${\bf T}_l$'s account for an in-plane translation of the corresponding
Bravais lattice with respect to a reference layer.  In analogy with bulk 2D systems [see Supplemental Material], we consider both triangular and unidirectional charge density order, reminiscent of the charge density wave states observed on the surface of the transition metal dichalcogenide NbSe$_2$ \cite{sou13}, and restrict, for simplicity, to the corresponding simplest sets of in-plane wavevectors of equal magnitude $G_0$. The long-range part of the free energy functional Eq.~\ref{eq:f0longrange} is then minimized   for in-plane charge density modulations oriented along the same direction but shifted by a value that is commensurate with their period $2 \pi / G_0$. Specifically, sinusoidal charge density waves are shifted by half of their period in adjacent layers (${\bf T}_l =  \left\{ \pi l / G_0 , 0 \right\}$) whereas in-plane charge density modulations with triangular symmetry render a three-layer stacking periodic structure  with the vector ${\bf T}_l = \left\{ 4 \pi l / (3 G_0), 0 \right\} $ [see Supplemental Material]. 
The allowed ${\bf G}$'s and the amplitudes $\phi_{{\bf G} , l}$ become independent of the layers index $l$, {\it i.e.} $\phi_{{\bf G} , l} = \phi_{{\bf G}}$. 
Precisely as in isotropic 2D and 3D systems \cite{ort08}, the optimal wavevector magnitude $G_0$  is  selected by the competition among the Coulomb interaction and the gradient squared term in the total free energy functional ${\cal F}_{\phi}$. 

\begin{figure}
\includegraphics[width=\columnwidth]{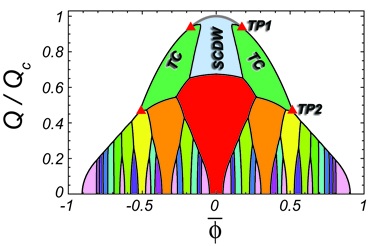}
\caption{Phase diagram for layer spacing $a_3= 5 \, \xi$. As before, the strength of the Coulomb interaction has been measured in units of the maximum coupling strength $Q_c = 0.5999...$. The phase diagram displays triple points  from in-plane unidirectional to triangular modulated structures   (TP1) and additional triple points (TP2) marking the disappearance of in-plane charge modulations.}
\label{fig:fig3}
\end{figure}

Having these combined in-plane and out-of-plane charge density textures in our hands, we have then determined the phase diagram in the $Q - \bar{\phi}$ plane by continuously changing the interlayer spacing $a_3$ while taking concomitantly into account the electronically charged superlattices. 
As anticipated above, for small interlayer distances ($a_3 < \sqrt{6}$) [see Supplemental Material] in-plane charge density modulations are completely suppressed and the phase diagram of Fig.~\ref{fig:fig1b} holds. 
Increasing $a_3$ 
a region of  in-plane unidirectional sinusoidal charge density modulations (SCDW) first appears in the phase diagram at strong coupling [c.f. Fig.~\ref{fig:fig2}]. The SCDW's amplitude 
vanishes along a Gaussian instability line $Q_g^{SCDW}=Q_c \left(a_3\right) f(\bar{\phi}, Q, a_3)$ where $Q_c \left(a_3 \right)$ indicates 
the maximum coupling strength above which only homogeneous states are allowed. 
As $Q$ is lowered, we find
this second-order phase transition line to converge to the 1/1
instability line $Q_{G}^{1/1}$.
This, in turn, implies the appearance of two $a_3$-dependent Lifshitz
points $\left\{ \bar{\phi}_L, Q_L \right\} = \left\{ \pm \sqrt{1/ 3 -
    2 /a_3^2} , \sqrt{24 / (\pi a_3^3)} \right\} $  where the period
of the SCDWs diverges and their  amplitude vanishes [see
Fig.\ref{fig:fig2} and Supplemental Material]. 
The presence of a second-order transition line to in-plane
unidirectional charge density modulations is unique of anisotropic 
systems. Indeed it is not encountered in isotropic systems both in 2D [Supplementary Information] and 3D \cite{ort08}. Such a feature is relevant for cuprates, where charge density wave ordering has been experimentally observed \cite{cha12,ghi12,com14,net14}, as it allows for a second-order charge density wave quantum critical point relevant for superconductivity \cite{cas95}.

The occurrence of the Lifshitz points is lost as  $a_3$ is increased further. 
The in-plane Gaussian instability line $Q_g^{SCDW}$ indeed crosses 
a first-order transition line leading to in-plane triangular lattices of inhomogeneities at two new triple points 
$ \left\{ \pm \bar{\phi}_{TP1} , Q_{TP1}  \right\}$ (with $Q_{TP1} >Q_L$) [see Fig.\ref{fig:fig3}] which are found to be exponentially close to the critical point of the phase diagram $\left\{ \bar{\phi} =0 , Q_c(a_3) \right\}$ in the $a_3 \gg 1$ regime [Supplementary Information]. 
Two additional triple points ($TP2$ in Fig.\ref{fig:fig3}) 
at lower coupling strength mark the disappearance of in-plane charge density modulations. 
Independent of the $a_3$ value, the appearance of electronically charged superlattices is thus in general preserved 
at weak coupling [c.f. Figs.~\ref{fig:fig2},\ref{fig:fig3}].

We have shown, in conclusion, that unidirectional superlattices of electronic charge can be spontaneously formed in layered materials with electronic phase separation tendencies. These phase segregated states realize an electronic structure analogous to the artificial unidirectional superlattices built in conventional semiconductors  by a periodic variation of composition
during epitaxial growth \cite{esa70} or in heterostructures of oxides \cite{oht02}. The electronic superlattice period depends on the average charge and, within each period made of $p+q$ layers,  the ratio $p/q$ among the number of crystalline planes in the minority $p$ and majority $q$  electronic phases is
governed by Farey fractions. 
By additionally varying the strength of Coulomb frustration -- the ratio between the energy cost due to long-range forces and the typical phase separation energy gain --  the phase diagram from the homogeneous phase to the inhomogeneous one displays Arnold tongues \cite{sch05} with the largest tongues, ordered by size, appearing according to a  modified Farey tree construction.

A long-period ($p+q\sim 10$)  spontaneous electronic
heterostructure, compatible with our proposal and consisting of
alternating antiferromagnetic and superconducting planes, has been
reported in iron-selenide superconductors\cite{char12,tex12}. A
short-period one ($p+q= 2$) has been observed on LuFe$_2$O$_4$ with
bilayers playing the role of basic units. 
These observations suggest that this phenomenon might be more common
than previously thought. Thus effects typical of heterostructures like
Wannier-Stark quantization\cite{ble88} and non-linear transport\cite{wac02} 
may appear in these layered charge-ordered correlated systems with
interesting perspectives for applications.

{\it Acknowledgements} --   C.O. acknowledges support from FP7-NMP-2011-EU-Japan project (No.283204 SUPER-IRON). 
J.L. acknowledges support from the Italian Institute of Technology through the project NEWDFESCM.

\end{document}